\documentclass[amssymb,aps,floatfix,showkeys,showpacs]{revtex4}
\usepackage{graphicx}
\usepackage{fancyvrb}
\usepackage{amsmath}

\begin{document}

\title{Corrigendum to ``Universal factorization of $3n-j$ ($j>2$) symbols\ldots'' [J. Phys. A: Math. Gen. 37 (2004) 3259]}

\author{Richard J. Mathar} 
\email{mathar@strw.leidenuniv.nl}
\homepage{www.strw.leidenuniv.nl/~mathar}
\affiliation{Leiden Observatory, Leiden University, P.O. Box 9513, 2300 RA Leiden, The Netherlands}

\pacs{03.65.Fd, 02.30.Gp, 02.20.Qs}

\date{\today}
\keywords{Angular momentum coupling, Wigner 12-j symbol, 3n-j symbol, Python}

\begin{abstract}
Ten values of 12$-j$ symbols of the first kind published earlier
are challenged
by values calculated with an independent Python program.
The program first implements a narrow
class of square roots of rational numbers, utilizing Python's
unlimited representation of big integers.
Wigner's $3jm$ symbols,
6-$j$, 9-$j$, 12-$j$ and 15-$j$ symbols
are then calculated
by their familiar representations as sums over products of these.
\end{abstract}

\maketitle 

\section{Update on 12-$j$ symbols}

As remarked earlier 
\cite{AndersonJCP129,RagniIJQC110}, the table of 12$-j$ symbols of the first kind
listed by Wei and Dalgarno \cite{WeiJPhysA37,WeiArxiv0306} needs correction.
Modified reference values are 
proposed in Table \ref{tab.12j}, generated with the Python package delivered
in the ancillary files. Each parameter set of angular momenta
is fed into an instance of the \texttt{Wigner12j} class, the instance is called,
and its value and floating point approximation are printed.
Appendix \ref{app.ex12j} provides one manual check of the simplest case.

Table \ref{tab.12j2} prints reference values
for the 12$-j$ symbols of the \emph{second} kind,
computed for a randomly selected set of half-integer input vectors.

\begin{table}
\caption{Reference values of 12-$j$ symbols of the first kind.}
\begin{ruledtabular}
\begin{tabular}{lll}

$\left\{\begin{array}{cccccccc}
j_1 & & j_2 & & j_3 & & j_4 \\
& l_1 & & l_2 & & l_3 & & l_4 \\
k_1 & & k_2 & & k_3 &&  k_4 \\
\end{array}\right\}$ & exact representation& decimal approximation \\
\hline

$\left\{\begin{array}{cccccccc}
0 & & 1 & & 1 & & 1 \\
& 1 & & 1 & & 0 & & 1 \\
1 & & 1 & & 1 & & 1 \\
\end{array}\right\}$ &
\verb+1/54+
&
0.0185185185185
\\

$\left\{\begin{array}{cccccccc}
1 & & 1 & & 2 & & 2 \\
& 2 & & 2 & & 2 & & 1 \\
1 & & 2 & & 1 & & 2 \\
\end{array}\right\}$ &
\verb+-7/3000*(7/3)^(1/2)+
&
-0.00356422554052
\\

$\left\{\begin{array}{cccccccc}
2 & & 1 & & 2 & & 1 \\
& 2 & & 2 & & 1 & & 2 \\
1 & & 1 & & 2 & & 2 \\
\end{array}\right\}$ &
\verb+1/600*(7/3)^(1/2)+
&
0.00254587538609
\\

$\left\{\begin{array}{cccccccc}
3 & & 3 & & 5 & & 4 \\
& 2 & & 4 & & 5 & & 3 \\
4 & & 2 & & 4 & & 5 \\
\end{array}\right\}$ &
\verb+-1025/3841992*(5/13)^(1/2)+
&
-0.000165455319732
\\

$\left\{\begin{array}{cccccccc}
4 & & 3 & & 5 & & 6 \\
& 3 & & 5 & & 6 & & 4 \\
2 & & 4 & & 3 & & 5 \\
\end{array}\right\}$ &
\verb+1/143143*(17/11)^(1/2)+
&
8.68476363508e-06
\\

$\left\{\begin{array}{cccccccc}
5 & & 5 & & 4 & & 3 \\
& 6 & & 4 & & 6 & & 4 \\
2 & & 4 & & 7 & & 5 \\
\end{array}\right\}$ &
\verb+-4457/4969107*(19/5005)^(1/2)+
&
-5.5263583797e-05
\\

$\left\{\begin{array}{cccccccc}
6 & & 4 & & 7 & & 4 \\
& 2 &&  5 & & 7 & & 6 \\
4 & & 5 & & 6 & & 3 \\
\end{array}\right\}$ &
\verb+40201/1032470010*(19/130)^(1/2)+
&
1.48855379744e-05
\\

$\left\{\begin{array}{cccccccc}
7 & & 8 & & 9 & & 10 \\
& 8 & & 6 & & 4 & & 6 \\
7 & & 9 & & 7 & & 5 \\
\end{array}\right\}$ &
\verb+-167145847027/25854768127188*(1/285285)^(1/2)+
&
-1.21036254096e-05
\\

$\left\{\begin{array}{cccccccc}
10 & & 7 & & 8 & & 6 \\
& 9 & & 10 & & 6 & & 8 \\
7 & & 9 & & 10 & & 7 \\
\end{array}\right\}$ &
\verb+438509/135635524675*(957/910)^(1/2)+
&
3.31543358353e-06
\\

$\left\{\begin{array}{cccccccc}
20 & & 15 & & 9 & & 10 \\
& 14 & & 18 & & 15 & & 15 \\
9 & & 8 & & 10 & & 12 \\
\end{array}\right\}$ &
\verb+-28068059458324/13772930246561475*(2/1431494295)^(1/2)+
&
-7.61739062208e-08
\\

\end{tabular}
\end{ruledtabular}
\label{tab.12j}
\end{table}

\begin{table}
\caption{Reference values of 12-$j$ symbols of the second kind.}
\begin{ruledtabular}
\begin{tabular}{lll}

$\left[\begin{array}{cccccccc}
j_1 &  j_2  & j_3  & j_4 \\
l_1  & l_2  & l_3  & l_4 \\
k_1 &  k_2 &  k_3 &  k_4 \\
\end{array}\right]$ & exact representation& decimal approximation \\
\hline
$\left[\begin{array}{cccc}
6 & 4 & 7 & 4 \\
6 & 7 & 4 & 4 \\
2 & 5 & 7 & 1 \\
\end{array} \right]$ &
\verb+ -28/23595*(1/195)^(1/2) + &
-8.49807860695e-05 \\

$\left[\begin{array}{cccc}
6 & 4 & 7 & 4 \\
6 & 7 & 4 & 5 \\
2 & 5 & 7 & 1 \\
\end{array} \right]$ &
\verb+ -7/47190*(119/2145)^(1/2) + &
-3.49387927903e-05 \\

$\left[\begin{array}{cccc}
6 & 4 & 7 & 4 \\
6 & 7 & 5 & 3 \\
2 & 5 & 7 & 1 \\
\end{array} \right]$ &
\verb+ 4/212355*(14/39)^(1/2) + &
1.12857185282e-05 \\

$\left[\begin{array}{cccc}
5.5 & 4.5 & 6.5 & 3.5 \\
6 & 6 & 5 & 3 \\
1.5 & 4.5 & 6.5 & 0.5 \\
\end{array} \right]$ &
\verb+ 1/572572*(57)^(1/2) + &
1.3185825425e-05 \\

$\left[\begin{array}{cccc}
5.5 & 4.5 & 6.5 & 3.5 \\
6 & 6 & 5 & 4 \\
1.5 & 4.5 & 6.5 & 0.5 \\
\end{array} \right]$ &
\verb+ 1/40898*(19/14)^(1/2) + &
2.84846384914e-05 \\

$\left[\begin{array}{cccc}
5.5 & 4.5 & 6.5 & 3.5 \\
6 & 6 & 6 & 3 \\
1.5 & 4.5 & 6.5 & 0.5 \\
\end{array} \right]$ &
\verb+ -3/572572*(19/91)^(1/2) + &
-2.39412737646e-06 \\

$\left[\begin{array}{cccc}
10.5 & 4.5 & 11.5 & 3.5 \\
4.5 & 16 & 4.5 & 16 \\
8 & 7 & 14.5 & 19.5 \\
\end{array} \right]$ &
\verb+ -73/878560*(14147/21487898)^(1/2) + &
-2.13199726694e-06 \\

$\left[\begin{array}{cccc}
10.5 & 4.5 & 11.5 & 3.5 \\
15 & 12 & 15 & 11 \\
12.5 & 12.5 & 11.5 & 12.5 \\
\end{array} \right]$ &
\verb+ -41530704637/111423970458157500*(629/19)^(1/2) + &
-2.14456489187e-06 \\

$\left[\begin{array}{cccc}
10.5 & 4.5 & 11.5 & 3.5 \\
11.5 & 9 & 14.5 & 3 \\
16 & 16 & 5.5 & 4.5 \\
\end{array} \right]$ &
\verb+ -15680464081/328436652258000*(7/221)^(1/2) + &
-8.49689356966e-06 \\

\end{tabular}
\end{ruledtabular}
\label{tab.12j2}
\end{table}

\section{Python3 Program}
\subsection{Auxiliary class of square roots}
The programming language Python uses an internal representation of integers
with unlimited precision, and offers a representation of rational numbers
on that basis in the \texttt{fractions.Fraction} class.

The module \texttt{surd} reproduced in
the auxiliary files
defines a \texttt{surd.Surd} class which represents a product of such a 
rational number by a positive square root of another such number, which suffices
to calculate the Wigner symbols ``exactly'' \cite{DeveikisPPNL4}. Floating
point representations are calculated on demand calling the \texttt{float}
or \texttt{to\_decimal} member functions. Multiplication and division
of two \texttt{Surd} are forwarded to the \texttt{Fraction}
implementation.

The task of keeping the number under the square root square-free is
delegated to functionality provided through the Python package
\texttt{NZMATH} \cite{nzmath}, see
\texttt{README.txt} in the auxiliary files.

Heuristically, the summations always reduce to members of a single quadratic field;
sums (or differences) of square roots of rational
numbers are apparently not needed \cite{StoneCPC21}.
(For the 9-$j$ symbols, this representability is a result of Wu's factorizations \cite{WuJMP13}.)
As a backup, the class \texttt{surd.SurdVec}
defines an exact representation of values of this kind,
stored as vectors of the type \texttt{surd.Surd},
and also bestowed with the basic arithmetic binary functions.

\subsection{Formulas Implemented}
The following standard representations are implemented in
the \texttt{wigner3j} module
\cite{Yutsis,RaschSIAMSC25,BalcarSTMP234}.
The class \texttt{Wigner3jm} calculates one value of
\begin{eqnarray}
&&
\left(
\begin{array}{ccc}
j_1& j_2 & j\\
m_1& m_2 & m\\
\end{array}
\right)
=\delta_{m_1+m_2,-m}
(-1)^{j_1-j_2-m}
\sqrt{
\frac{(j+j_1-j_2)!(j-j_1+j_2)!(j_1+j_2-j)!
}{(j_1+j_2+j+1)!}
}
\nonumber \\
&&
\times
\sqrt{
(j+m)!(j-m)!(j_1-m_1)!(j_1+m_1)!(j_2-m_2)!(j_2+m_2)!
}
\nonumber \\
&&
\times
\sum_{k} 
\frac{(-1)^k}{k!(j_1+j_2-j-k)!(j_1-m_1-k)!
(j_2+m_2-k)!(j-j_2+m_1+k)!
(j-j_1-m_2+k)!
}
.
\end{eqnarray}
The class \texttt{Wigner6j} uses
\begin{eqnarray}
&&
\left\{
\begin{array}{ccc}
j_1& j_2 & j_3\\
l_1& l_2 & l_3\\
\end{array}
\right\}
=\Delta(j_1,j_2,j_3)
\Delta(j_1,l_2,l_3)
\Delta(l_1,j_2,l_3)
\Delta(l_1,l_2,j_3)
\nonumber \\
&&
\times
\sum_z (-1)^z\frac{(z+1)!}{(z-j_1-j_2-j_3)!
(z-j_1-l_2-l_3)!
(z-l_1-j_2-l_3)!
(z-l_1-l_2-j_3)!
}
\nonumber\\
&&
\times
\frac{1}{(j_1+j_2+l_1+l_2-z)!
(j_1+j_3+l_1+l_3-z)!
(j_2+j_3+l_2+l_3-z)!
}
,
\label{eq.6j}
\end{eqnarray}
with triangular factors defined as
\begin{equation}
\Delta(j_1,j_2,j_3)\equiv
\left[\frac{(j_1+j_2-j_3)!(j_1-j_2+j_3)!(-j_1+j_2+j_3)!}{(j_1+j_2+j_3+1)!}\right]^{1/2}
.
\end{equation}
The class \texttt{Wigner9j} implements
\begin{equation}
\left\{
\begin{array}{ccc}
j_1 & j_2 & j_3\\
l_1 & l_2 & l_3\\
k_1 & k_2 & k_3\\
\end{array}
\right\}
=
\sum_x (2x+1)(-1)^{2x}
\left\{
\begin{array}{ccc}
j_1 & j_2 & j_3\\
l_3 & k_3 & x\\
\end{array}
\right\}
\left\{
\begin{array}{ccc}
l_1 & l_2 & l_3\\
j_2 & x & k_2\\
\end{array}
\right\}
\left\{
\begin{array}{ccc}
k_1 & k_2 & k_3\\
x & j_1 & l_1\\
\end{array}
\right\}
.
\end{equation}
The results of this 9-$j$ calculation have been validated against other
published values
\cite{RaoCPC56,LaiCPC70,WeiCPC120,WeiCP12}.
Other values, without the factor $(-)^{2x}$, have also appeared \cite{RahmanCJM63}.
In the class \texttt{Wigner12j},
the 12$-j$ symbols of the first kind are \cite{JahnPR93,OrdPR94,JangJMP9,AdamsPRA24}
\begin{equation}
\left\{
\begin{array}{cccccccc}
j_1 && j_2 && j_3 && j_4\\
&l_1 && l_2 && l_3 && l_4\\
k_1 && k_2 && k_3 && k_4\\
\end{array}
\right\}
=
\sum_x (2x+1)(-1)^{R_n-x}
\left\{
\begin{array}{ccc}
j_1 & k_1 & x\\
k_2 & j_2 & l_1\\
\end{array}
\right\}
\left\{
\begin{array}{ccc}
j_2 & k_2 & x\\
k_3 & j_3 & l_2\\
\end{array}
\right\}
\left\{
\begin{array}{ccc}
j_3 & k_3 & x\\
k_4 & j_4 & l_3\\
\end{array}
\right\}
\left\{
\begin{array}{ccc}
j_4 & k_4 & x\\
j_1 & k_1 & l_4\\
\end{array}
\right\},
\label{eq.12jI}
\end{equation}
where $R_n\equiv \sum_{i=1}^n (j_i+l_i+k_i)$ at $n=4$.

The symmetric 12$-j$  symbols
of the second kind are \cite{Yutsis,SauJPG8}
\begin{equation}
\left[
\begin{array}{cccc}
j_1 & j_2 & j_3 & j_4\\
l_1 & l_2 & l_3 & l_4\\
k_1 & k_2 & k_3 & k_4\\
\end{array}
\right]
=
(-1)^{l_1-l_2-l_3+l_4}
\sum_x (2x+1)
\left\{
\begin{array}{ccc}
k_1 & k_2 & x\\
j_3 & j_1 & l_1\\
\end{array}
\right\}
\left\{
\begin{array}{ccc}
k_3 & k_4 & x\\
j_3 & j_1 & l_2\\
\end{array}
\right\}
\left\{
\begin{array}{ccc}
k_1 & k_2 & x\\
j_4 & j_2 & l_3\\
\end{array}
\right\}
\left\{
\begin{array}{ccc}
k_3 & k_4 & x\\
j_4 & j_2 & l_4\\
\end{array}
\right\}
.
\end{equation}
Representative output of six evaluations with the program is gathered in
Table \ref{tab.12j2}.

In the classes \texttt{Wigner15j} and \texttt{Wigner18j}, the symbols of the first and second kind are
\cite[(17.1),(17.2)]{Yutsis}
\begin{equation}
\left\{
\begin{array}{cccccccc}
j_1 && j_2 && \ldots && j_n\\
&l_1 && l_2 && \ldots && l_n\\
k_1 && k_2 && \ldots && k_n\\
\end{array}
\right\}
=
\sum_x (2x+1)(-1)^{R_n+(n-1)x}
\left\{
\begin{array}{ccc}
j_n & k_n & x\\
j_1 & k_1 & l_n\\
\end{array}
\right\}
\prod_{i=1}^{n-1}
\left\{
\begin{array}{ccc}
j_i & k_i & x\\
k_{i+1} & j_{i+1} & l_i\\
\end{array}
\right\}
\end{equation}
and
\begin{equation}
\left[
\begin{array}{cccccccc}
j_1 && j_2 && \ldots && j_n\\
&l_1 && l_2 && \ldots && l_n\\
k_1 && k_2 && \ldots && k_n\\
\end{array}
\right]
=
\sum_x (2x+1)(-1)^{R_n+nx}
\left\{
\begin{array}{ccc}
j_n & k_n & x\\
k_1 & j_1 & l_n\\
\end{array}
\right\}
\prod_{i=1}^{n-1}
\left\{
\begin{array}{ccc}
j_i & k_i & x\\
k_{i+1} & j_{i+1} & l_i\\
\end{array}
\right\}
\end{equation}
at $n=5$ and $n=6$, respectively.
Tables \ref{tab.15j1} and \ref{tab.15j2} show numerical results of these.

\begin{table}
\caption{Reference values of 15-$j$ symbols of the first kind.}
\begin{ruledtabular}
\begin{tabular}{lll}

& exact representation& decimal approximation \\
\hline
$\left\{\begin{array}{cccccccccc}
3 && 3 && 2 && 4 && 2.5 \\
& 4 && 4 && 5 && 2.5 && 5 \\
2.5 && 2.5 && 4.5 && 4.5 && 3 \\
\end{array} \right\}$ &
\verb! -16939/2904545952*(13/55)^(1/2) ! &
-2.83530541752e-06 \\

$\left\{\begin{array}{cccccccccc}
2 && 4.5 && 5 && 4.5 && 4 \\
& 2.5 && 3.5 && 4.5 && 2.5 && 4.5 \\
3.5 && 2 && 2.5 && 4 && 3.5 \\
\end{array} \right\}$ &
\verb! -161153/4840909920*(1/429)^(1/2) ! &
-1.60724745689e-06 \\

$\left\{\begin{array}{cccccccccc}
4.5 && 3 && 3 && 5 && 4.5 \\
& 3.5 && 1 && 5 && 1.5 && 4.5 \\
2 && 2.5 && 2.5 && 2.5 && 1 \\
\end{array} \right\}$ &
\verb! -13/1829520*(13/6)^(1/2) ! &
-1.04592908944e-05 \\

$\left\{\begin{array}{cccccccccc}
4.5 && 2.5 && 4.5 && 5 && 2.5 \\
& 2 && 4 && 3.5 && 3.5 && 3 \\
3.5 && 3.5 && 4.5 && 3 && 1.5 \\
\end{array} \right\}$ &
\verb! 13451/65197440*(1/231)^(1/2) ! &
1.35743186372e-05 \\

$\left\{\begin{array}{cccccccccc}
4 && 1 && 2.5 && 5 && 4.5 \\
& 4 && 2.5 && 2.5 && 3.5 && 4.5 \\
5 && 5 && 3.5 && 4 && 2.5 \\
\end{array} \right\}$ &
\verb! 323/13172544*(1/39)^(1/2) ! &
3.92645442513e-06 \\

$\left\{\begin{array}{cccccccccc}
0.5 && 3 && 3.5 && 5 && 4 \\
& 2.5 && 4.5 && 3.5 && 3 && 3.5 \\
2.5 && 2 && 4.5 && 5 && 3 \\
\end{array} \right\}$ &
\verb! 147899/512265600*(1/429)^(1/2) ! &
1.39393139182e-05 \\

\end{tabular}
\end{ruledtabular}
\label{tab.15j1}
\end{table}

\begin{table}
\caption{Reference values of 15-$j$ symbols of the second kind.}
\begin{ruledtabular}
\begin{tabular}{lll}

& exact representation& decimal approximation \\
\hline
$\left[\begin{array}{cccccccccc}
3.5 && 3.5 && 5 && 0.5 && 2 \\
& 6 && 6.5 && 5.5 && 2.5 && 5.5 \\
5 && 4 && 5.5 && 6 && 6.5 \\
\end{array} \right]$ &
\verb! 346789/9717364800*(119/4290)^(1/2) ! &
5.94376514774e-06 \\

$\left[\begin{array}{cccccccccc}
4 && 1.5 && 2 && 3 && 3.5 \\
& 3.5 && 2.5 && 4 && 2.5 && 0.5 \\
4 && 1.5 && 1 && 3 && 3.5 \\
\end{array} \right]$ &
\verb! -1157/3386880*(1/35)^(1/2) ! &
-5.7743024419e-05 \\

$\left[\begin{array}{cccccccccc}
2.5 && 1.5 && 1 && 2.5 && 3.5 \\
& 2 && 0.5 && 1.5 && 2 && 4 \\
5 && 4 && 4.5 && 5 && 5 \\
\end{array} \right]$ &
\verb! 47/118800*(1/231)^(1/2) ! &
2.6030075474e-05 \\

$\left[\begin{array}{cccccccccc}
4.5 && 2.5 && 4.5 && 5 && 2.5 \\
& 2 && 4 && 3.5 && 3.5 && 3 \\
3.5 && 3.5 && 4.5 && 3 && 1.5 \\
\end{array} \right]$ &
\verb! 10327/143434368*(1/42)^(1/2) ! &
1.11095459009e-05 \\

$\left[\begin{array}{cccccccccc}
4 && 1 && 2.5 && 5 && 4.5 \\
& 4 && 2.5 && 2.5 && 3.5 && 4.5 \\
5 && 5 && 3.5 && 4 && 2.5 \\
\end{array} \right]$ &
\verb! 8545/807288768*(1/7)^(1/2) ! &
4.00068296486e-06 \\

$\left[\begin{array}{cccccccccc}
0.5 && 3 && 3.5 && 5 && 4 \\
& 2.5 && 4.5 && 3.5 && 3 && 3.5 \\
2.5 && 2 && 4.5 && 5 && 3 \\
\end{array} \right]$ &
\verb! -2909/8781696*(1/1365)^(1/2) ! &
-8.96600541253e-06 \\

$\left[\begin{array}{cccccccccc}
3.5 && 3 && 4.5 && 3.5 && 4.5 \\
& 0.5 && 4.5 && 3 && 3 && 1 \\
3.5 && 3 && 4.5 && 4.5 && 2.5 \\
\end{array} \right]$ &
\verb! 50741/717171840*(1/455)^(1/2) ! &
3.31688256263e-06 \\

\end{tabular}
\end{ruledtabular}
\label{tab.15j2}
\end{table}

In the class \texttt{Wigner15j}, the symbols of the third kind are
\cite[(20.3)]{Yutsis}
\begin{equation}
\left\{
\begin{array}{ccccccccc}
k_1 && k_1' & k && k' & k_2 && k_2'\\
&p_1 &&& p &&& p_2 \\
j_1 && j_1' & j && j' & j_2 && j_2'\\
\end{array}
\right\}
=
\sum_x (2x+1)(-1)^{x+p-j-k'}
\left\{
\begin{array}{ccc}
k & j & x\\
j' & k' & p\\
\end{array}
\right\}
\left\{
\begin{array}{ccc}
k & j & x\\
k_1 & j_1 & p_1\\
k_2 & j_2 & p_2\\
\end{array}
\right\}
\left\{
\begin{array}{ccc}
k' & j' & x\\
k_1' & j_1' & p_1\\
k_2' & j_2' & p_2\\
\end{array}
\right\}
,
\end{equation}
generating Table \ref{tab.15j3}.

\begin{table}
\caption{Reference values of 15-$j$ symbols of the third kind.}
\begin{ruledtabular}
\begin{tabular}{lll}

& exact representation& decimal approximation \\
\hline
$\left\{\begin{array}{cccccccccc}
6.5 && 3.5 & 6 && 3.5 & 1.5 && 3 \\
& 3 &&& 4.5 &&& 4.5 \\
3.5 && 4.5 & 6.5 && 5 & 5 && 2.5 \\
\end{array} \right\}$ &
\verb! 75443/19434729600*(17/33)^(1/2) ! &
2.78617209378e-06 \\

$\left\{\begin{array}{cccccccccc}
5.5 && 6.5 & 5.5 && 5.5 & 3 && 4 \\
& 3.5 &&& 6 &&& 2 \\
5 && 6 & 5 && 3 & 1 && 6 \\
\end{array} \right\}$ &
\verb! -16529/2763375615*(34/77)^(1/2) ! &
-3.97466705158e-06 \\

$\left\{\begin{array}{cccccccccc}
2.5 && 1.5 & 6.5 && 3.5 & 6 && 5 \\
& 5.5 &&& 6 &&& 6.5 \\
3 && 6 & 3.5 && 6.5 & 3.5 && 2.5 \\
\end{array} \right\}$ &
\verb! 27441/2821634816*(5/1547)^(1/2) ! &
5.5289030385e-07 \\

$\left\{\begin{array}{cccccccccc}
5 && 3 & 5 && 2.5 & 2 && 2.5 \\
& 3 &&& 4.5 &&& 5 \\
4 && 1 & 2 && 2.5 & 5 && 3.5 \\
\end{array} \right\}$ &
\verb! 19267/8324316000*(1/2)^(1/2) ! &
1.6366301271e-06 \\

$\left\{\begin{array}{cccccccccc}
3.5 && 3.5 & 5 && 3.5 & 3.5 && 2 \\
& 4.5 &&& 1.5 &&& 2.5 \\
2 && 5 & 3 && 1.5 & 3 && 4.5 \\
\end{array} \right\}$ &
\verb! 211/1724800*(3/77)^(1/2) ! &
2.41467661705e-05 \\

$\left\{\begin{array}{cccccccccc}
4 && 4.5 & 4 && 1.5 & 1 && 4 \\
& 4 &&& 4.5 &&& 4 \\
4 && 1.5 & 5 && 3.5 & 4 && 4 \\
\end{array} \right\}$ &
\verb! -26003/146779776*(1/2310)^(1/2) ! &
-3.68596532572e-06 \\

\end{tabular}
\end{ruledtabular}
\label{tab.15j3}
\end{table}
The fourth kind is
\begin{eqnarray}
\left\{
\begin{array}{ccccccccc}
& j_1 && k_1 & s_1 & k_1' && j_1'\\
p && l && s && l' && p'\\
& j_2 && k_2 & s_2 & k_2' && j_2'\\
\end{array}
\right\}
=
(-1)^{k_1+k_2-s_1-s_2+p+p'+2l'}\sum_x (2x+1)
\nonumber \\ \times
\left\{
\begin{array}{ccc}
j_1 & j_2' & x\\
l & s_2 & j_2\\
s_1 & l' & j_1'\\
\end{array}
\right\}
\left\{
\begin{array}{ccc}
j_1 & j_2' & x\\
k_2' & k_1 & p\\
\end{array}
\right\}
\left\{
\begin{array}{ccc}
j_2 & j_1' & x\\
k_1' & k_2 & p'\\
\end{array}
\right\}
\left\{
\begin{array}{ccc}
k_1 & k_2' & x\\
k_2 & k_1' & s\\
\end{array}
\right\}
,
\label{eq.15j4}
\end{eqnarray}
which leads to Table \ref{tab.15j4}\@.

\begin{table}
\caption{Reference values of 15-$j$ symbols of the fourth kind.}
\begin{ruledtabular}
\begin{tabular}{lp{8cm}l}

& exact representation& decimal approximation \\
\hline
$\left\{\begin{array}{cccccccccc}
& 6&& 2& 0.5 & 6&& 5\\
4&& 5.5 && 5&& 5.5 && 3\\
& 2.5 && 4.5 & 3& 1.5 && 4.5 \\
\end{array} \right\}$ &
\verb! 101/17249760*(7/143)^(1/2) ! &
1.29544601101e-06 \\

$\left\{\begin{array}{cccccccccc}
& 6.5 && 5.5 & 4& 6.5 && 6.5 \\
6&& 3.5 && 3&& 4.5 && 6\\
& 5&& 2& 6.5 & 2&& 4\\
\end{array} \right\}$ &
\verb! -33049/4210858080*(7/429)^(1/2) ! &
-1.0025547351e-06 \\

$\left\{\begin{array}{cccccccccc}
& 4&& 5& 6.5 & 4&& 6\\
6&& 5.5 && 5&& 3.5 && 6\\
& 5.5 && 1.5 & 3& 5.5 && 6.5 \\
\end{array} \right\}$ &
\verb! -110945249/1354301649805104*(95)^(1/2) ! &
-7.98463566586e-07 \\

$\left\{\begin{array}{cccccccccc}
& 2.5 && 2.5 & 2& 4&& 3.5 \\
1&& 1.5 && 2.5 && 3.5 && 3.5 \\
& 4&& 4.5 & 3.5 & 4&& 1\\
\end{array} \right\}$ &
\verb!! &
0\\

$\left\{\begin{array}{cccccccccc}
& 4&& 1.5 & 1.5 & 2.5 && 2\\
2.5 && 4.5 && 4&& 1.5 && 0.5 \\
& 1.5 && 1& 3& 4&& 3.5 \\
\end{array} \right\}$ &
\verb! -5/54432*(1/14)^(1/2) ! &
-2.45500111986e-05 \\

$\left\{\begin{array}{cccccccccc}
& 2&& 2& 2.5 & 2&& 2\\
4&& 0.5 && 4&& 3.5 && 3\\
& 5&& 3& 4.5 & 4&& 2\\
\end{array} \right\}$ &
\verb! 139/370440*(1/165)^(1/2) ! &
2.92115735985e-05 \\

\end{tabular}
\end{ruledtabular}
\label{tab.15j4}
\end{table}
The fifth kind and Table \ref{tab.15j5} are based on \cite[(20.9)]{Yutsis}
\begin{eqnarray}
\left\{
\begin{array}{ccccc}
k_1 & k_1' & j_1 & l_1 & l_1'\\
k_2 & k_2' & j_2 & l_2 & l_2'\\
k_3 & k_3' & j_3 & l_3 & l_3'\\
\end{array}
\right\}
=
\sum_{x_1,x_2} (2x_1+1)(2x_2+1)(-1)^{j_1+j_2+k_1+k_1'-k_2+k_2'-l_2'+l_3+x_2}
\nonumber \\ \times
\left\{
\begin{array}{ccc}
l_2 & l_3' & x_1\\
k_2' & k_3' & k_1\\
\end{array}
\right\}
\left\{
\begin{array}{ccc}
l_2' & l_3 & x_2\\
k_2 & k_3 & k_1'\\
\end{array}
\right\}
\left\{
\begin{array}{ccc}
l_2 & l_3' & x_1\\
l_2' & l_3 & x_2\\
j_2 & j_3 & j_1\\
\end{array}
\right\}
\left\{
\begin{array}{ccc}
k_2' & k_3' & x_1\\
k_3 & k_2 & x_2\\
l_1 & l_1' & j_1\\
\end{array}
\right\}
.
\end{eqnarray}

\begin{table}
\caption{Reference values of 15-$j$ symbols of the fifth kind.}
\begin{ruledtabular}
\begin{tabular}{lp{8cm}l}

& exact representation& decimal approximation \\
\hline
$\left\{\begin{array}{cccccccccc}
1 & 4.5 & 3.5 & 5 & 4.5 \\
6.5 & 6 & 3.5 & 3 & 5.5 \\
5 & 2 & 4 & 5 & 5 \\
\end{array} \right\}$ &
\verb! -1430315/13264202952*(1/462)^(1/2) ! &
-5.01683187528e-06 \\

$\left\{\begin{array}{cccccccccc}
0.5 & 2.5 & 3 & 5.5 & 4.5 \\
2 & 2 & 6 & 2 & 5 \\
6.5 & 2.5 & 3 & 3.5 & 1.5 \\
\end{array} \right\}$ &
\verb! 17/288288*(1/2145)^(1/2) ! &
1.27323540877e-06 \\

$\left\{\begin{array}{cccccccccc}
4.5 & 5 & 5 & 4 & 6 \\
3 & 4.5 & 6 & 6.5 & 0.5 \\
5.5 & 4 & 4 & 4 & 1 \\
\end{array} \right\}$ &
\verb! -18433/218640708*(1/390)^(1/2) ! &
-4.26906428456e-06 \\

$\left\{\begin{array}{cccccccccc}
5 & 6.5 & 3.5 & 6 & 2.5 \\
1 & 3.5 & 4.5 & 5.5 & 5 \\
2.5 & 2.5 & 6 & 5.5 & 2.5 \\
\end{array} \right\}$ &
\verb! -19583/1205836632*(5/42)^(1/2) ! &
-5.6033933163e-06 \\

$\left\{\begin{array}{cccccccccc}
6 & 3 & 6 & 6 & 2 \\
3 & 6.5 & 3.5 & 1 & 2.5 \\
1.5 & 5 & 3.5 & 2 & 5.5 \\
\end{array} \right\}$ &
\verb! -103/4530240*(323/273)^(1/2) ! &
-2.47306739633e-05 \\

$\left\{\begin{array}{cccccccccc}
1.5 & 3 & 1.5 & 5 & 4.5 \\
3.5 & 4 & 3.5 & 3.5 & 4 \\
2 & 3 & 4 & 1.5 & 3.5 \\
\end{array} \right\}$ &
\verb! 30607/1173553920*(1/5)^(1/2) ! &
1.16636025701e-05 \\

$\left\{\begin{array}{cccccccccc}
5 & 2.5 & 4.5 & 4 & 2.5 \\
2 & 4 & 1 & 3.5 & 4.5 \\
2 & 1.5 & 3.5 & 4.5 & 3 \\
\end{array} \right\}$ &
\verb! -27397/1746360000*(13/231)^(1/2) ! &
-3.72164479622e-06 \\

\end{tabular}
\end{ruledtabular}
\label{tab.15j5}
\end{table}

\section{Summary}
A list of 10 values of the 12-$j$ symbols of the first kind that appeared
earlier in the literature has been corrected. Reference values for 12-$j$
symbols of both kinds and for 15-$j$ values of
all five kinds have been computed with a program written in Python3, which
is made available as ancillary material.

\bibliographystyle{apsrmp}
\bibliography{all}

\appendix

\section{Validation Example} \label{app.ex12j}
The verification of the first entry in Table \ref{tab.12j} may be
done with pen and paper. Eq.\ (\ref{eq.12jI}) is
\begin{equation}
\left\{
\begin{array}{cccccccc}
0 && 1 && 1 && 1\\
&1 && 1 && 0 && 1\\
1 && 1 && 1 && 1\\
\end{array}
\right\}
=
\sum_x (2x+1)(-1)^{12-x}
\left\{
\begin{array}{ccc}
0 & 1 & x\\
1 & 1 & 1\\
\end{array}
\right\}
\left\{
\begin{array}{ccc}
1 & 1 & x\\
1 & 1 & 1\\
\end{array}
\right\}
\left\{
\begin{array}{ccc}
1 & 1 & x\\
1 & 1 & 0\\
\end{array}
\right\}
\left\{
\begin{array}{ccc}
1 & 1 & x\\
0 & 1 & 1\\
\end{array}
\right\}
.
\end{equation}
The factor $\Delta(j_1,j_2,j_3)$ on the right hand side of (\ref{eq.6j})
applied to the first 6-$j$ symbol on the right hand side implies that
nonzero contributions may only emerge from $x$ equal to $-1$, $0$ or 1\@.
The special value \cite[9.5.1]{Varshalovich}
\begin{equation}
\left\{
\begin{array}{ccc}
0 & b & c\\
d & e & f\\
\end{array}
\right\}
= (-1)^{b+e+d}\frac{\delta_{bc}\delta_{ef}}{\sqrt{(2b+1)(2e+1)}}
\label{eq.6ja0}
\end{equation}
sets that first factor to zero if $x=-1$ or $x=0$. Only $x=1$ might contribute:
\begin{equation}
\left\{
\begin{array}{cccccccc}
0 && 1 && 1 && 1\\
&1 && 1 && 0 && 1\\
1 && 1 && 1 && 1\\
\end{array}
\right\}
=
3\times (-1)^{12-1}
\left\{
\begin{array}{ccc}
0 & 1 & 1\\
1 & 1 & 1\\
\end{array}
\right\}
\left\{
\begin{array}{ccc}
1 & 1 & 1\\
1 & 1 & 1\\
\end{array}
\right\}
\left\{
\begin{array}{ccc}
1 & 1 & 1\\
1 & 1 & 0\\
\end{array}
\right\}
\left\{
\begin{array}{ccc}
1 & 1 & 1\\
0 & 1 & 1\\
\end{array}
\right\}
.
\label{eq.tab1expl}
\end{equation}
From (\ref{eq.6ja0})
by symmetry of the 6$-j$ symbol \cite{Varshalovich}
\begin{equation}
\left\{
\begin{array}{ccc}
a & b & c\\
d & e & 0\\
\end{array}
\right\}
= (-1)^{a+b+c}\frac{\delta_{ae}\delta_{bd}}{\sqrt{(2a+1)(2b+1)}}
\end{equation}
we conclude that three factors on the right hand side are
\begin{equation}
\left\{
\begin{array}{ccc}
0 & 1 & 1\\
1 & 1 & 1\\
\end{array}
\right\}
=
\left\{
\begin{array}{ccc}
1 & 1 & 1\\
1 & 1 & 0\\
\end{array}
\right\}
=
\left\{
\begin{array}{ccc}
1 & 1 & 1\\
0 & 1 & 1\\
\end{array}
\right\}
=-1/3.
\label{eq.onethird}
\end{equation}
Furthermore \cite[Table 9.2]{Varshalovich}
\begin{equation}
\left\{
\begin{array}{ccc}
a & b & c\\
1 & c & b\\
\end{array}
\right\}
=
(-1)^{a+b+c+1}\frac{1}{2}\cdot\frac{-a(a+1)+b(b+1)+c(c+1)}{\sqrt{b(2b+1)(b+1)c(2c+1)(c+1)}}
\end{equation}
leads to
\begin{equation}
\left\{
\begin{array}{ccc}
1 & 1 & 1\\
1 & 1 & 1\\
\end{array}
\right\}
=
(-1)^4\frac{1}{2}\times\frac{-2+2+2}{\sqrt{3\times 2\times 3\times 2}}
= 1/6
.
\end{equation}
Inserting this value and (\ref{eq.onethird}) into (\ref{eq.tab1expl}) generates
the value $1/54$ on the top of Table \ref{tab.12j}.

\section{Book Errata}
Corrigenda to the translation of the book by Yutsis \textit{et al}
\cite{Yutsis} are:
\begin{itemize}
\item
Our equation (\ref{eq.15j4}) corrects typographic errors
(concerning primes on the right hand sides) in \cite[(20.6)]{Yutsis} and also in
\cite[(9)]{ElbatanoniNP82}.
\item
All directions of the lines $l_1$, $l_2$,\ldots $l_n$ for the symbols of the
first kind in \cite[Fig 17.1]{Yutsis}
should be reversed, as already reported \cite{MattisPRD39}.
\item
Immediately related to the previous bullet, the directions of the lines
$l_1$,\ldots, $l_4$ for the 12-$j$ symbols of the first kind
in \cite[Fig. 19.1a]{Yutsis} and \cite[Fig. 19.1b]{Yutsis} need to be reversed.
(The five sign changes by the inversion of the five $l$
for the 15-$j$ symbols of the first kind in \cite[Fig. 20.2a-b]{Yutsis}, however,
appear to be already compensated by the $+-$-``handedness'' in
these figures.)
\end{itemize}

\end{document}